\documentclass[aps,prd,preprint,groupedaddress,showpacs]{revtex4}

\usepackage{epsfig}
\usepackage{graphics}

\newcommand{\tr}{\mathop{\rm tr}\nolimits}
\newcommand{\cE}{{\cal E}}
\newcommand{\cL}{{\cal L}}
\newcommand{\pa}{\partial}
\newcommand{\ii}{{\rm i}}
\newcommand{\vp}{\varphi}
\begin{document}

\title{Skyrmions and Faddeev-Hopf Solitons}

\author{R. S. Ward}
\email[]{richard.ward@durham.ac.uk}
\affiliation{Department of Mathematical Sciences, University of
Durham, Durham DH1 3LE}

\date{\today}

\begin{abstract}
This paper describes a natural one-parameter family of generalized Skyrme
systems, which includes the usual SU(2) Skyrme model and the Skyrme-Faddeev
system.  Ordinary Skyrmions resemble polyhedral shells, whereas the Hopf-type
solutions of the Skyrme-Faddeev model look like closed loops, possibly linked
or knotted.  By looking at the minimal-energy solutions in various topological
classes, and for various values of the parameter, we see how the polyhedral
Skyrmions deform into loop-like Hopf Skyrmions.
\end{abstract}


\maketitle

%%%%%%%%%%%%%%%%%%%%%%%%%%%%%%%%%%%%%%%%%%%%%%%%%%%%%%%%%%%%%%%%%%%%%

\section{Introduction}

Recent years have seen extensive progress on understanding the nature and
dynamics of topological solitons \cite{MS04}, and in particular of Skyrmions.
For the SU(2) Skyrme system, minimal-energy Skyrmions resemble polyhedral
shells \cite{BS02}; for example, the 3-Skyrmion looks like a tetrahedron
\cite{BTC90}.  On the other hand, the Hopf-type solitons in the Skyrme-Faddeev
system (where the field takes values in the 2-sphere $S^2$) resemble closed
loops, which may be linked or knotted \cite{BS99}; for example, the 3-soliton
in this system looks like a slightly-twisted circular loop.  This paper
describes a natural one-parameter family of generalized Skyrme systems, which
interpolates between the standard SU(2) Skyrme model and the Skyrme-Faddeev
model.  Its minimum-energy solutions interpolate between polyhedral Skyrmions
and string-like Hopf solitons.

The simplest way to describe the family is as follows.  In the SU(2) Skyrme model,
the field takes values in the 3-sphere $S^3$, with its standard metric.  This
3-sphere is fibred over $S^2$ (the Hopf fibration); and instead of the standard
metric on $S^3$, we can use a metric for which distances along the
(one-dimensional) fibres are scaled by a factor which we denote $1-\alpha$.
So $\alpha=0$ gives the standard Skyrme system, whereas $\alpha=1$ corresponds
to the target space being the quotient $S^2$, namely the Skyrme-Faddeev system.
The global symmetry SO(4) in the $\alpha=0$ case is broken to U(2) when
$\alpha>0$; and this in turn means that the generalized Skyrmion solutions for
$\alpha>0$ have less symmetry than those for $\alpha=0$.

The system can also be formulated in terms of a pair $Z=(Z^1,Z^2)^t$ of complex
scalars, and as such is related to condensed-matter
systems in which there are two flavours of Cooper pairs \cite{BFN02}.
The parameter $\alpha$ then appears, in particular, as the coefficient of a
term $J_{\mu} J^{\mu}$, where $J_{\mu}=\ii Z^\dagger\,\pa_{\mu}Z$ is the current
density.  

In two spatial dimensions, and without the fourth-order Skyrme terms, the case
$\alpha=1$ corresponds to the $CP^1$ model. The generalization of this to
$\alpha<1$ was investigated in \cite{Z86}.  It arises as a modification of the
$CP^1$ model which takes account of the effect of fermions (starting
with a system which has fermions as well as bosons, and integrating out the
fermionic degrees of freedom).  In this case, there are explicit finite-energy
static solutions (parametrized by $\alpha$) which, for $\alpha=1$, are the
usual instanton solutions of the 2-dimensional $CP^1$ model.
In the three-dimensional case which is discussed below, one needs a Skyrme term
to stabilize the solutions; and the solutions have to be obtained numerically.

%%%%%%%%%%%%%%%%%%%%%%%%%%%%%%%%%%%%%%%%%%%%%%%%%%%%%%%%%%%%

\section{Family of Skyrme Systems}

Let us consider, first, the general situation of a map $\Phi$ from a
3-space (with local coordinates $x^j$ and metric $g_{jk}$) to another
3-space (with local coordinates $\vp^a$ and metric $H_{ab}$).  The Skyrme
energy density $\cE$ of such a map may be defined as follows \cite{M87},
in terms of the differential $\pa_j\vp^a$ of $\Phi$. Define a $3\times3$
matrix $D$ by
\begin{equation} \label{D}
  {D_a}^b = g^{jk} (\pa_j\vp^c) H_{ac} (\pa_k\vp^b).
\end{equation}
Then $\cE=\cE_2+\cE_4$, where
\begin{equation} \label{genE}
  \cE_2 = \lambda_2\,\tr(D), \quad
     \cE_4 = \frac{1}{2}\lambda_4\left[(\tr D)^2-\tr(D^2)\right].
\end{equation}
Here $\lambda_2$ and $\lambda_4$ are constants.  If the metric $H_{ab}$
admits a group of symmetries (isometries), then these will correspond to
(global) symmetries of the system.

In what follows, we take the target space to be the 3-sphere $S^3$ equipped
with a one-parameter family of U(2)-invariant metrics.  A particular member
of this family is the standard SO(4)-invariant metric, and the corresponding
system is the usual SU(2) Skyrme model.  The family of metrics
may be described as follows.

Let $Z=(Z^1,Z^2)^t$ denote a complex 2-vector satisfying the constraint
$Z^\dagger\,Z = |Z^1|^2+|Z^2|^2 = 1$ (where $Z^\dagger$ is the
complex-conjugate row vector corresponding to the column vector $Z$).
The set of all such vectors $Z$ forms a 3-sphere.  Note that the
map $(Z^1,Z^2)\mapsto Z^1/Z^2$ is the standard Hopf fibration from $S^3$
to $S^2$, with $Z^1/Z^2$ being the usual stereographic coordinate on $S^2$.
The standard metric $G$ on $S^3$ corresponds to
\begin{equation} \label{StMet}
  ds^2=dZ^\dagger\,dZ.
\end{equation}
Let $\xi$ be the vector field obtained from the 1-form
$\omega=-\ii Z^\dagger\,dZ$ by raising its index with the metric (\ref{StMet}).
This vector field $\xi$ has unit length, and is tangent to the fibres of the
Hopf fibration.  Our family of metrics, parametrized by the real number
$\alpha$, is taken to be $H=G-\alpha\,\omega\otimes\omega$.
An alternative way to write $H$ is
\begin{equation} \label{Met}
  ds^2=dZ^\dagger\,dZ + \alpha(Z^\dagger\,dZ)(Z^\dagger\,dZ).
\end{equation}
Note that both $G$ and $\omega$, and hence also $H$, are manifestly invariant
under the U(2) transformations $Z\mapsto \Lambda Z$, where
$\Lambda\in{\rm U(2)}$.

For $\alpha<1$, the metric (\ref{Met}) is positive-definite.  But when
$\alpha=1$, it becomes degenerate, with $\xi$ being a zero-eigenvector;
distances along the Hopf fibres are then zero, and the metric is, in effect,
the standard metric on the quotient space $CP^1\cong S^2$.
In other words, our one-parameter family includes the standard 3-sphere
($\alpha=0$) and the standard 2-sphere ($\alpha=1$).  We will restrict to the range
$\alpha\leq1$ for which the metric is non-negative; in fact, our interest is
in the range $0\leq\alpha\leq1$, which interpolates between the Skyrme and
the Skyrme-Faddeev systems.

The Lagrangian $\cL$ of the generalized Skyrme system (consistent with the
expressions (\ref{genE}) for the static energy density) may be described
as follows.  The vector $Z$ determines an SU(2) matrix according to
\[
  U = \left[\begin{array}{cc}
            Z^1 & -\bar Z^2\\
            Z^2 &  \bar Z^1
         \end{array}\right].
\]
Write
\[    %\begin{eqnarray*}
 U^\dagger\pa_{\mu}U = L_{\mu} = \ii L^a_{\mu}\sigma_a, \quad
 [L_{\mu},L_{\nu}] = K_{\mu\nu} = \ii K^a_{\mu\nu}\sigma_a,
\]
where the partial derivative is with respect to space-time coordinates
$x^\mu$, and where $\sigma_a$ denotes the Pauli matrices.  Then
$\cL=\cL_2+\cL_4$, where
\begin{eqnarray}
 \cL_2 &=& \lambda_2\,g^{\mu\nu}(L^a_\mu L^a_\nu-\alpha L^3_\mu L^3_\nu),\\
 \cL_4 &=& \frac{1}{8}\lambda_4\,g^{\mu\nu} g^{\beta\gamma}
     \left[(1-\alpha)K^a_{\mu\beta} K^a_{\nu\gamma}
       + \alpha K^3_{\mu\beta} K^3_{\nu\gamma}\right].
\end{eqnarray}
In this form, the global U(2) symmetry corresponds to
$U\mapsto \Omega U \Gamma$, where
$\Gamma$ is an SU(2) matrix and $\Omega = \exp(\ii\theta\sigma_3)$ is a diagonal
SU(2) matrix; note that this transformation preserves both $L^a_\mu L^a_\nu$
and $L^3_\mu$.

If $\alpha=0$, then $\cL$ is the standard Skyrme Lagrangian.  If $\alpha=1$,
on the other hand, we get the Skyrme-Faddeev system
\cite{F75,FN97,GH97,BS99,HS99,W99,W00,BW01}.  One way of seeing this is to replace
the field $Z$ by the unit 3-vector field
$\vec\psi=Z^\dagger \vec\sigma Z$.  Then $\cL$ with $\alpha=1$ becomes
\[
  \cL = \frac{1}{4}\lambda_2(\pa_{\mu}\vec\psi)^2
       + \frac{1}{32}\lambda_4 (\Omega_{\mu\nu})^2,
\]
where $\Omega_{\mu\nu}=
\vec\psi\cdot(\pa_{\mu}\vec\psi)\times(\pa_{\nu}\vec\psi)$; this is the
Skyrme-Faddeev Lagrangian.

If we take the space on which the field $U$ is defined to be ${\bf R}^3$,
then we need a boundary condition $U\to U_0$ (constant) as $r\to\infty$,
to have finite energy.  Fields satisfying this condition are 
classified topologically by their winding number $N=\int{\cal B}\,d^3x$,
where ${\cal B}$ is the topological charge density
\begin{equation} \label{B}
  {\cal B} = \epsilon_{jkl}\tr(L_j L_k L_l)/(24\pi^2).
\end{equation}
In the limit $\alpha\to1$, $N$ equals the Hopf number of the $S^2$-valued field.

The values of the constants $\lambda_2$ and $\lambda_4$ correspond to the energy
and length scales.  To choose convenient values for them
in what follows, let us consider the system defined on the unit 3-sphere
$S^3$ (that is, take $g_{jk}$ to be the standard metric on $S^3$)
\cite{M87,W99}; and take the field $Z(x^j)$ to correspond to the identity
map from $S^3$ to itself (in other words, an isometry if $\alpha=0$).
It is straightforward to compute the energy $E$ of this field: one gets
\[
  E = 2\pi^2(3-\alpha)\lambda_2 + 2\pi^2(3-2\alpha)\lambda_4.
\]
So from now on let us take
\[
  \lambda_2=1/[4\pi^2(3-\alpha)], \quad \lambda_4=1/[4\pi^2(3-2\alpha)]; 
\]
consequently, the `identity' field has unit energy for all $\alpha\in[0,1]$.

%%%%%%%%%%%%%%%%%%%%%%%%%%%%%%%%%%%%%%%%%%%%%%%%%%%
\section{Families of Skyrmion Solutions}

A numerical minimization procedure was used to find local minima of the static
energy $E$ for various values of $N$ and $\alpha$, and hence stable Skyrmion
solutions; the results are described below.
The procedure uses a finite-difference version of the functional
$E$ on a cubic grid, with a second-order scheme in which the truncation
error is of order $h^4$ where $h$ is the lattice spacing, and using the
coordinate $1/x$ for $|x|>q\approx1$ (similarly for $y$ and $z$) so that
the whole of ${\bf R}^3$ is included.  With a relatively small number of
lattice points (say $33^3$), this achieves an accuracy of better than $1\%$.
The discrete energy was then minimized using a standard conjugate-gradient
method (flowing down the energy gradient).  This produces a local minimum of
the energy functional.  In general, there are many local minima; the starting
configuration determines which one is produced by this procedure.  It seems
likely that the solutions described below are global minima in the relevant
topological classes, but the only evidence for this at present is consistency
with previous studies in the $\alpha=0$ and $\alpha=1$ cases
\cite{BTC90,BS99,BS02}.

Most straightforward are the $N=1$ and $N=2$ cases, where the solutions admit
a continuous symmetry.  For $N=1$, the $\alpha=0$ Skyrmion has O(3) (spherical)
symmetry, and energy $E=1.232$. When $\alpha>0$, this is broken to O(2) (axial)
symmetry.  The normalized energy $E(\alpha)$ depends smoothly on $\alpha$, and
the numerical results indicate that, to within the small numerical error,
its dependence is quadratic: $E(\alpha)=1.232-0.008\,\alpha^2$.  The topological
charge density (\ref{B}) has an almost-spherical shape, for all~$\alpha$.

For $N=2$, one has O(2) symmetry both for $\alpha=0$ and $\alpha=1$, and the
constant-${\cal B}$ surfaces resemble tori.  So the expectation is that the
$N=2$ generalized Skyrmions will look like tori for all values of $\alpha$,
with $E(\alpha)$ decreasing from $E(0)=2.358$ to $E(1)=2.00$ \cite{GH97,W00}
over the range $\alpha\in[0,1]$; but this has not been checked.

It is worth remarking at this point on the energy-values of Skyrme-Faddeev
solitons given in \cite{BS99}, so as to facilitate comparison with that paper.
The energies in \cite{BS99} should be divided by a factor of $32\pi^2\sqrt{2}$
in order to adjust the normalization to the one being used here; and by a further
factor of (about) $0.93$ to allow for the fact that \cite{BS99} used a finite-size
box (rather than all of ${\bf R}^3$).  For example, in the $N=2$ case, \cite{BS99}
gives an energy $E_{{\rm BS}}=835$, which when divided by the two factors
above yields $E=2.01$.  This is within $0.5\%$ of the correct figure.

\begin{figure}
  \includegraphics[scale=0.8]{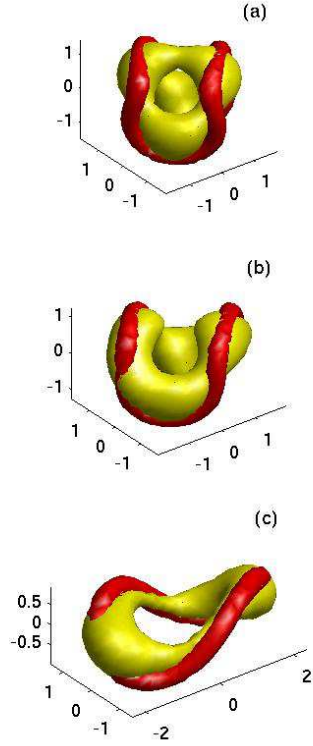}
  \caption{The charge density isosurface and position-curve $\psi_3=-0.8$
   of the $N=3$ generalized Skyrmions: for (a) $\alpha=0$, (b) $\alpha=0.2$, and
   (c) $\alpha=0.4$. \label{fig1}}
\end{figure}

For $N\geq3$, the picture is less straightforward, with the Skyrmions having
at most discrete symmetry.  We look in detail at the cases $N=3$
and $N=4$.  The 3-Skyrmion (for $\alpha=0$) has energy $E=3.4386$ and tetrahedral
symmetry \cite{BTC90,BS02}; in particular, a typical constant-${\cal B}$ surface
resembles a tetrahedron.  It is also useful to plot the curve in ${\bf R}^3$ where
$\psi_3=-1$, or equivalently where $Z^1=0$ and $|Z^2|=1$; in the Faddeev-Skyrme
system, this curve may be interpreted as the position of the string-like Hopf
Skyrmion \cite{BS99}.  Each plot in Figure 1 depicts the surface
${\cal B}({\bf x})=(\max{\cal B})/2$, with the `thickened' curve $\psi_3=-0.8$
strung around it; subfigure (a) is for $\alpha=0$, (b) is for $\alpha=0.2$,
and (c) is for $\alpha=0.4$.  We see that as $\alpha$ increases from zero,
the tetrahedral Skyrmion transforms into a twisted torus or loop (see also the
pictures in \cite{BS99} for the $\alpha=1$ case).  The tetrahedral symmetry
is broken to the subgroup $D_2$.  The normalized energy $E(\alpha)$ again
has a quadratic dependence on $\alpha$: $E(\alpha)\approx3.4386-0.60\,\alpha^2$.

\begin{figure}
  \includegraphics[scale=0.8]{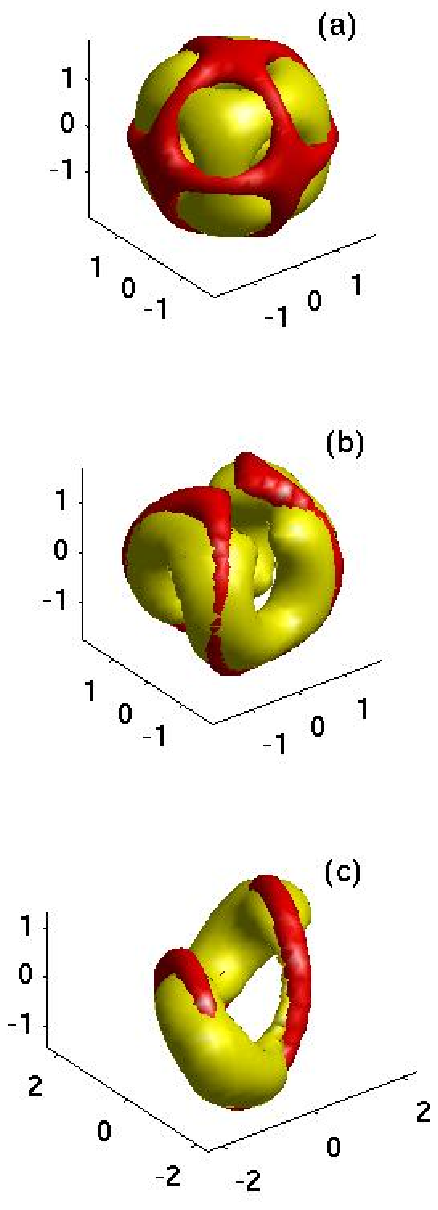}
  \caption{The charge density isosurface and position-curve $\psi_3=-0.8$ of
   the $N=4$ generalized Skyrmions: for (a) $\alpha=0$, (b) $\alpha=0.35$, and
   (c) $\alpha=0.4$. \label{fig2}}
\end{figure}

Finally, let us look at the case $N=4$.  The 4-Skyrmion (for $\alpha=0$) resembles
a cube \cite{BTC90,BS02}: see Figure 2(a), where the same quantities are plotted
as in Figure 1.  As $\alpha$ increases, the minimum-energy configuration becomes
a closed loop strung along eight edges of the cube (Figure 2(b), for $\alpha=0.35$),
which then flattens as $\alpha$ increases further.  When $\alpha=1$, one again
gets a twisted circular loop, with the twisting being greater than in the $N=3$
case (see also the pictures in \cite{BS99}).

We have seen that the Skyrme model and the Skyrme-Faddeev-Hopf system may be
regarded as members of a one-parameter family of generalized Skyrme systems;
and the topological-soliton solutions of all these systems, although rather
different in appearance, are all closely-related to one another.
A recent paper \cite{KS04} has pointed out a similarity between sphaleron
solutions of the Skyrme system and axially-symmetric Hopf solitons, especially
as the winding number $N$ increases.  These solutions are unstable
(saddle-points of their respective energy functionals); and this connection
between Skyrmions and Hopf solitons is quite different from the one described
above.  It may be of interest, however, to investigate sphaleron-type solutions
of the family of Skyrme systems, and see how they depend on the family parameter
$\alpha$.

\noindent{\bf Note added in proof.} A similar family arises from
considering bundles of strings [S. Nasir and A. J. Niemi,
Mod. Phys. Lett. A 17, 1445 (2002)].  The author is grateful
to Prof. Niemi for correspondence regarding this.

\begin{acknowledgments}
 Support from the UK Engineering and Physical Sciences Research Council
  is gratefully acknowledged.
\end{acknowledgments}


%\end{document}

\end{document}